\documentclass[aip,jmp,amsmath,amssymb, reprint]{revtex4-1}

\usepackage{amsmath,amssymb,graphicx,color,bm,epstopdf,float,dcolumn,bm,soul}

\begin{document}
\preprint{AIP/123-QED}

\title{Optical diode based on exciton-polaritons}

\author{T. Espinosa-Ortega}
\email{teortega@ntu.edu.sg}
\affiliation{Division of Physics and Applied Physics, Nanyang Technological University 637371, Singapore}

\author{T. C. H. Liew}
\affiliation{Division of Physics and Applied Physics, Nanyang Technological University 637371, Singapore}

\author{I. A. Shelykh}
\affiliation{Division of Physics and Applied Physics, Nanyang Technological University 637371, Singapore}
\affiliation{Science Institute, University of Iceland, Dunhagi-3, IS-107, Reykjavik, Iceland}

\date{\today}

%%%%%%%%%%%%%%%%%%%%%%%%%%%%%%%%%%%%%%%%

\begin{abstract}
We propose theoretically an optical diode based on exciton-polaritons in semiconductor microcavities. A flow of polaritons in the bistable regime is used to send signals through an asymmetric fixed potential that favours the tunneling of particles in one direction. Through dynamic modelling of the coherent polariton field, we demonstrate the characteristics of an ideal diode, namely that the forward signal is fully transmitted while the transmission in the reverse direction tends to zero, without any additional external control. Moreover, the system proves to be robust to the presence of disorder, intrinsic to microcavities, and can function at gigahertz repetition rates.
\end{abstract}
\maketitle
%%%%%%%%%%%%%%%%%%%%%%%%%%%%%%%%%%%%%%

The engineering of optical microprocessors would open the door for a new generation of computers, allowing fast calculations, where the low carrier resistance would reduce heat losses. The search for the optimal systems to develop this technology has increased the interest in non-linear optical controlled systems such as exciton-polaritons in semiconductor microcavities~\cite{Sanvitto2012,Carusotto2013}.

Polaritons are quasi-particles resulting from the strong coupling of photons and quantum well excitons. This origin has provided them with a low mass and strong nonlinear interactions -- characteristics that have made them successful building blocks for multiple photonic devices, such as lasers \cite{Chrisopoulos2007,Bajoni2008,KenaCohen2010,Schneider}, interferometers \cite{Ivan1,Sturm}, holographic devices \cite{Timo}, terahertz emitters~\cite{Kavokin2010,Kyriienko2013}, optical routers\cite{Hugo}, transistors\cite{Ivan2,Gao2012} and integrated photonic circuits \cite{Ballarini,Espinosa}, among others~\cite{Liew2011}. With such a variety of applications, it would seem that a new era of polaritonics is close.

A common problem for the linking of polaritonic devices is the compensation of loss due to photon emission. The amplification of polariton states by bosonic stimulation~\cite{Savvidis2000,Dima2000,Saba2001,Lagoudakis2002} was one of the first nonlinear effects studied in these systems and propagating polariton wavepackets were recently amplified through a non-resonant excitation of reservoir states~\cite{Wertz2012}. An alternative approach to amplification has been based on the bistable behaviour of polaritons~\cite{Whittaker,Baas,Gippius}, which itself has been used to create polariton switches~\cite{Adrados2011,DeGiorgi2012} and ultrafast memories~\cite{Cerna2013}. In the bistable regime, a switched state survives indefinitely~\cite{Cerna2013} where the loss is compensated by a continuous wave laser that, together with interactions, sets the bistability in the system. By exploiting the propagation of a domain wall, it has been predicted theoretically that one can transport loss-less signals in the microcavity plane\cite{Timo1,Timo2}. Structured potentials control the signal propagation along wires or ``polariton neurons'' and signals can cover distances that go far beyond the limits expected of the polariton lifetime.

To be competitive, a signal processing device needs to achieve a low error rate, which may require the suppression of feedback. A useful element to reduce feedback would be a polaritonic analogue of the diode -- a device that ideally allows full signal transmission in one direction, with signal suppression in the other. Very recently, a double barrier resonant diode based on polaritons was constructed\cite{Nguyen}, where a high contrast gating of the transmission was achieved through the application of a gating laser. This design was based on ballistically propagating polariton wavepackets\cite{Wertz2010}.

In this paper we present a design for an optical diode compatible with propagating domain walls (polariton neurons). An asymmetric response is achieved without using an additional gating laser, instead making use of a fixed spatially patterned potential, which favours the tunneling of particles in one direction only. We show that it is possible to achieve  $100\%$ transmission for the forward signal, while the backward signal is fully blocked by the static potential. Without the action of any external electric or optical control potential, the device can operate as a tunnel diode or as a backward diode, depending on the direction of the incoming flow of polaritons. Finally we test the functionality of the system when disorder in the microcavity is present and show that both signal propagation along polariton neurons and the diode itself is fully robust to realistic levels of disorder.
\begin{figure}[b!]
\includegraphics[width=5.cm,height=4.5cm]{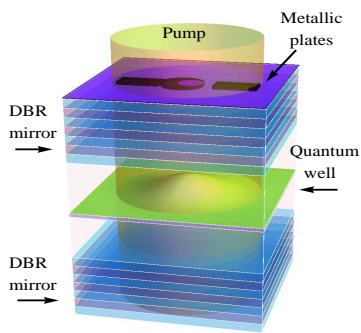}
\caption{A microcavity formed by two distributed Bragg mirrors (DBRs). The excitons in the quantum well will strongly couple to the photons injected by the normal incident optical pump, forming polaritons. The metallic plates in the top will provide the fixed potential.}
\label{Fig1}
\end{figure}

\begin{figure}[h!]
\includegraphics[width=0.95\linewidth]{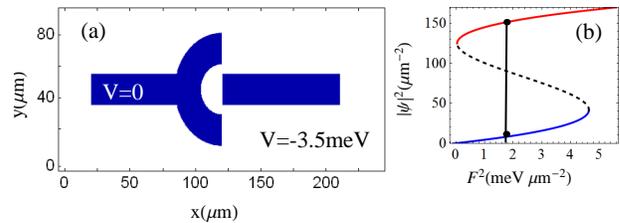}
\caption{(a) Potential $V(\textbf{r})$. (b) Eq.(\ref{bistability}), for $V(\textbf{r})=0$; the lower (blue) and the upper (red) branches are the possible states of the system,  the dashed line indicates an unstable state and the vertical line indicates the value of $F_{cw}^2$ chosen for our subsequent calculations.}
\label{Fig2}
\end{figure}

The system consists of an optical microcavity, shown in Fig.\ref{Fig1}, where the polaritons within are subjected to a fixed potential $V(\textbf{r})$, which takes the pattern shown in Fig.\ref{Fig2}(a). The potential patterning can be achieved by a variety of different techniques, such as: varying the thickness of the cavity layers \cite{Kaitouni}; etching the cavity structure\cite{Wertz2010}; using optically induced potentials~\cite{Amo,Tosi}; or placing metallic plates on the top of the microcavity\cite{Lai}, as shown in Fig.\ref{Fig1}.

In order to describe the dynamics of the system, we use the mean-field approximation that leads to the Gross-Pitaevskii equation \cite{Carusotto},
\begin{equation}
\label{GP}
i\hbar\frac{\partial\Psi}{\partial t}=\left(\widehat{E}_{LP}(\widehat{\textbf{k}})+V(\textbf{r})-\frac{i\hbar}{2\tau}+\alpha|\Psi|^2\right)\Psi+\mathfrak{F}(\textbf{r},t)e^{-\frac{i\varepsilon_{p}t}{\hbar}}.
\end{equation}
$\Psi$ is the coherent polariton field; $V(\textbf{r})$ is the spatial potential; $\tau$ is the polariton lifetime; $\alpha$ is the nonlinear polariton-polariton interaction strength; and $\mathfrak{F}(\textbf{r},t)$ is the incident optical field amplitude with energy $\varepsilon_{p}$. The kinetic energy corresponds to the non-parabolic lower polariton dispersion band $\widehat{E}_{LP}(\widehat{\textbf{k}})=\left(E_0(\widehat{\textbf{k}})+E_X-\sqrt{(E_0(\widehat{\textbf{k}})-E_X)^2+4\Omega^2}\right)/2$, where $E_0(\widehat{\textbf{k}})$ is the photon dispersion, $E_X$ is the exciton energy, and $\Omega$ is the exciton-photon coupling strength. For simplicity, we neglect the exciton dispersion, which is flat compared to the photon dispersion and shift the origin of energy to $E_{LP}(0)$. We also neglect the spin degree of freedom, noting that the interaction strength between polaritons with opposite spins is typically weaker than that between parallel spins~\cite{Ciuti1998}.

Before understanding how polaritons at different points in the microcavity are coupled, it is instructive to first consider the local response of the system. This can be done by first neglecting the kinetic energy term, in which an analytical expression relates the local polariton density of a stationary state to the optical pump intensity\cite{Whittaker}:
\begin{equation}
\label{bistability}
|\mathfrak{F}|^2=\left[\left(\varepsilon_p-V(\textbf{r})-\alpha|\Psi|^2\right)^2+\frac{\hbar^2}{4\tau^2}\right]|\Psi|^2.
\end{equation}
When $(\varepsilon_{p}-V(\textbf{r}))^2>3\hbar/(2\tau)$, the above equation describes an ``S" shaped response curve, and the polariton density can take either of the values given by the upper or lower branch (the middle branch is unstable\cite{Whittaker}), as shown in Fig.\ref{Fig2}(b). Therefore, by varying the laser intensity it is possible to drive the system into a bistable regime. Now, when the kinetic energy is taken into account, a coupling of the dynamic response of spatially separated domains allows the sending of signals, as will be described briefly.

The microcavity will be subjected to both a continuous wave (cw) optical pump $F_{cw}(\textbf{r})$ and a pulsed excitation $\mathfrak{P}(\textbf{r},t)$, both with Gaussian shape and the same energy. The total field is then given by the superposition $\mathfrak{F}(\textbf{r},t)=F_{cw}(\textbf{r}) +\mathfrak{P}(\textbf{r},t)$. The optical pump $F_{cw}(\textbf{r})$, will provide the whole area with a constant wave to compensate the finite lifetime of polaritons; the intensity $F_{cw}^2$ can be increased steadily until the value marked by the black vertical line in Fig.\ref{Fig2}(b), so the initial state will be given by the black dot in the lower bistable branch.

The pulse $\mathfrak{P}(\textbf{r},t)$ will be used to send the signals. It will locally increase the intensity of the optical field, which will cause the polaritons in the lower bistable branch to ``jump" to the upper branch. After a time $\delta t$, the pulse $\mathfrak{P}(\textbf{r},t)$ will vanish and the polaritons remaining in the upper branch will relax to the corresponding intensity $F_{cw}^2$ [black dot in the upper branch in Fig.\ref{Fig2}(b)]. The kinetic energy term then couples neighbouring points in space, such that neighbouring polaritons will jump successively to the upper bistable branch\cite{Shelykh2008}, allowing signal propagation in the form of a propagating domain wall\cite{Timo1}.

Notice that the optical pump will excite polaritons mainly in the blue region of Fig.\ref{Fig2}(a) where $V(\textbf{r})=0$. This is due to the energy detuning of the pumps with respect to the potential energy, which will be smaller in this region, compared to the detuning for polaritons in the area where $V(\textbf{r})=-3.5meV$ [see Eq.(\ref{bistability})]. This choice of the potential, instead of a confinement well, is done to avoid the interference of reflected waves.
\begin{figure}[h!]
\includegraphics[width=1\linewidth]{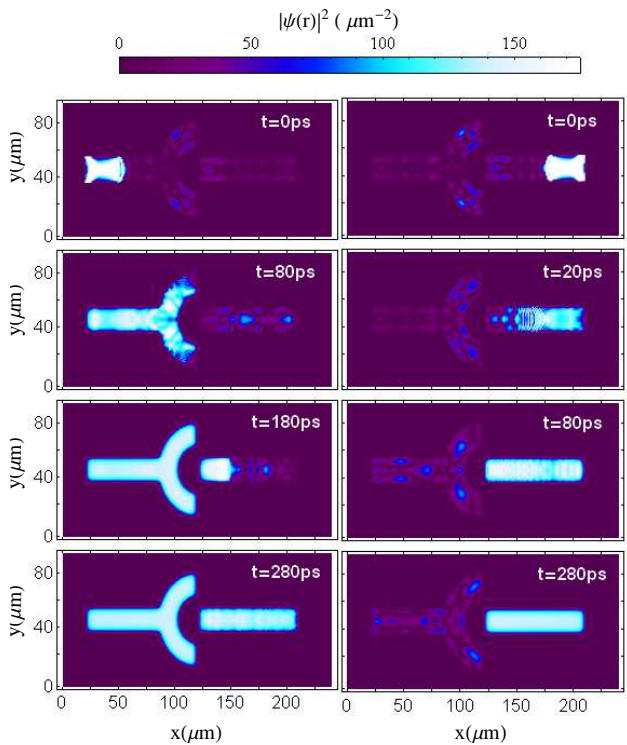}
\caption{In the first column, we show the forward signal; in the subsequent rows we appreciate the transition of the signal beyond the semicircle due to constructive interference. The second column corresponds to the backward signal. In this case the signal can't propagate to the left side. The parameters used are, $\tau=18ps$, $\Omega=2.5meV$, $\varepsilon_p=0.5meV$, $\alpha=0.004meV\mu m^2$. The photon dispersion was taken as parabolic with an effective mass $10^{-5}$ of the free electron mass.}
\label{Fig3}
\end{figure}
%
%%%%%%%%%%%%%%%%%%%%%%%%%%%%%%%%%%%%%%%%%%%%%%%%%%%%%%%%%%%

First we consider the case of \textit{forward operation}, where the signal originates from the extreme left in Fig.\ref{Fig2}(a). In the first column of Fig.\ref{Fig3} we show the numerical results of Eq.(\ref{GP}) at different times. In the first row we show the moment when the pulse $\mathfrak{P}(\textbf{r},t)$ is sent. The slides below show the system evolving after $\mathfrak{P}(\textbf{r},t)$ has vanished. When the signal arrives at the ``C" looking shape of the potential, polaritons begin to ballistically propagate from all points on the semicircle across the gap. They constructively interfere in the center of the semicircle, giving a sufficient intensity to switch the state of the second part of the channel. The signal then continues propagating to the right, as shown in Fig.\ref{Fig3}. The transmitted signal has the same intensity as the incident one, corresponding to $100\%$ transmission, since optical losses are fully compensated by the cw pump.

%%%%%%%%%%%%%%%%%%%%%%%%%%%%%%%%%%%%%%%%%%%%%%%%%%%%%%%%%

The \textit{reverse operation} corresponds to the case where the signal starts in the extreme right (second column in Fig.\ref{Fig3}). The mechanism for signal propagation is completely analogous to the previous case, nevertheless, here the signal can't propagate beyond the semicircle. This is partly because the smaller size of the edge of the right-hand channel causes fewer polaritons to be emitted. More importantly, any polaritons that are emitted and attempt to cross the gap in the reverse direction are not focused, in contrast to the forward operation case. While a few polaritons can cross the gap, their intensity is not enough to switch the signal in the semicircle region. In other words, the potential's shape completely blocks the reverse signal. Note that this asymmetric response does not rely on any additional external electric or optical field, as in traditional tunneling diodes, but merely in the direction of the incoming flow of polaritons.
\begin{figure}[t!]
\includegraphics[width=0.7\linewidth]{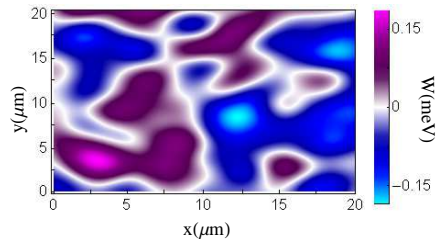}
\caption{We show a small area of the disorder potential Eq.(\ref{W}). The parameters characteristic of the disorder are $\sigma=1.5\mu m$ and $A_w=0.052meV$.}
\label{Fig4}
\end{figure}
%
%%%%%%%%%%%%%%%%%%%%%%%%%%%%%%%%%%%%%%%%%%5
%
\begin{figure}[b!]
\includegraphics[width=1\linewidth]{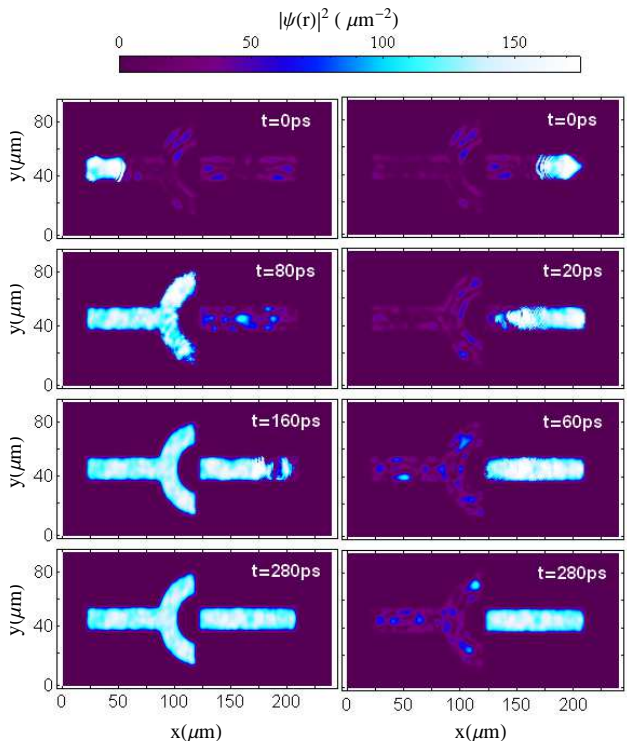}
\caption{The same as Fig.\ref{Fig3}, but in the presence of the random potential Eq.(\ref{W}).}
\label{Fig5}
\end{figure}

Disorder is unavoidable in all realistic microcavity structures~\cite{Savona} and represents a potential feedback mechanism given that it causes the Rayleigh scattering of polaritons~\cite{Langbein2002} in all directions and enhanced backscattering~\cite{Gurioli2005}. Consequently, we tested the robustness of our system when defects are present in the potential. We add to the potential $V(\textbf{r})$ Fig.\ref{Fig2}(a), a random potential of the form,
\begin{equation}
W(\textbf{r})=\frac{A_w}{\sqrt{\pi}\sigma}\int_0^\infty Exp\left[-\frac{(\textbf{r-r'})^2}{2\sigma^2}\right]v(\textbf{r})d\textbf{r'},
\label{W}
\end{equation}
with $v(\textbf{r})$ a random white noise potential with uncorrelated points, $\sigma$ the correlation length, and $A_w$ the root mean squared (rms) fluctuation (see Fig.\ref{Fig4}).

The numerical calculations, show that the diode functionality is not compromised for a realistic disorder with rms fluctuation of $0.052meV$ (Fig.\ref{Fig5}). For a hypothetical disorder with much stronger fluctuations, the transmission of the signals is eventually affected and the diode loses its asymmetric response.

Finally, it is worth to remark that the power demand on the system is determined by the optical pump that injects polaritons in the bistable regime; it has been experimentally achieved\cite{Paraiso} with a power consumption of $50Wcm^2$, although it could be improved by several orders of magnitude, by increasing the lifetime of polaritons\cite{Espinosa}.

In conclusion, we have demonstrated theoretically how an optical diode based on exciton-polariton bistability can be constructed. Numerical calculations prove that a forward propagating signal is fully transmitted, while the transmission for a reverse propagating signal tends to zero. Moreover, the asymmetric response is inherent to a fixed potential and the system proves to be robust in the presence of disorder. These characteristics make our design a promising candidate for future applications.

\begin{acknowledgments}
The work was supported by
Tier1 project ``Novel polariton devices" and FP7 IRSES project POLAPHEN
\end{acknowledgments}

%%%%%%%%%%%%%%%%%%%%%%%%%%%%%%%%%%%%%%%

\end{document}